  \documentstyle[12pt,procsla,epsfig]{article}
  
  \catcode`\@=11
  \long\def\@makefntext#1{
  \protect\noindent \hbox to 3.2pt {\hskip-.9pt  
  $^{{\ninerm\@thefnmark}}$\hfil}#1\hfill}		
  
  \def\@makefnmark{\hbox to 0pt{$^{\@thefnmark}$\hss}}  
  	
  \def\ps@myheadings{\let\@mkboth\@gobbletwo
  \def\@oddhead{\hbox{}
  \rightmark\hfil\ninerm\thepage}   
  \def\@oddfoot{}\def\@evenhead{\ninerm\thepage\hfil
  \leftmark\hbox{}}\def\@evenfoot{}
  \def\sectionmark##1{}\def\subsectionmark##1{}}
  
\begin{document}
  
  \centerline{\normalsize\bf THE LAKE BAIKAL EXPERIMENT: SELECTED RESULTS}
  \baselineskip=16pt
  
  \vfill
  \vspace*{0.6cm}
\noindent 
{\footnotesize 
{V.A.BALKANOV$^a$,I.A.BELOLAPTIKOV$^g$, L.B.BEZRUKOV$^a$, N.M.BUDNEV$^b$, A.G.CHENSKY$^b$, \\
I.A.DANILCHENKO$^a$, Zh.-A.M.DZHILKIBAEV$^a$, G.V.DOMOGATSKY$^a$, A.A.DOROSHENKO$^a$, \\
S.V.FIALKOVSKY$^d$, O.N.GAPONENKO$^a$, A.A.GARUS$^a$, T.I.GRESS$^b$, D.KISS$^i$, A.M.KLABUKOV$^a$, \\
A.I.KLIMOV$^f$, S.I.KLIMUSHIN$^a$, A.P.KOSHECHKIN$^a$, Vy.E.KUZNETZOV$^a$, V.F.KULEPOV$^d$, \\
L.A.KUZMICHEV$^c$, S.V.LOVZOV$^b$, B.K.LUBSANDORZHIEV$^a$, M.B.MILENIN$^d$, R.R.MIRGAZOV$^b$, \\
N.I.MOSEIKO$^c$, V.A.NETIKOV$^a$, E.A.OSIPOVA$^c$, A.I.PANFILOV$^a$, Yu.V.PARFENOV$^b$, \\
A.A.PAVLOV$^b$, E.N.PLISKOVSKY$^a$, P.G.POHIL$^a$, E.G.POPOVA$^c$, M.I.ROZANOV$^e$, V.Yu.RUBZOV$^b$, \\
I.A.SOKALSKY$^a$, CH.SPIERING$^h$, O.STREICHER$^h$, B.A.TARASHANSKY$^b$, G.TOHT$^i$, T.THON$^h$, \\
R.VASILJEV$^a$, R.WISCHNEWSKI$^h$, I.V.YASHIN$^c$.}}

\vspace{3mm}
{\footnotesize\it
$^a$\ Institute for Nuclear Research, Moscow, Russia 

$^b$\ Irkutsk State University, Irkutsk, Russia 

$^c$\ Institute of Nuclear Physics, MSU, Moscow, Russia 

$^d$\ Nizhni  Novgorod  State  Technical University, Nizhni  Novgorod, Russia 

$^e$\ St.Petersburg State  Marine Technical  University, St.Petersburg, Russia 

$^f$\  Kurchatov Institute, Moscow, Russia 

$^g$\ Joint Institute for Nuclear Research, Dubna, Russia 

$^h$\ DESY-Zeuthen, Berlin/Zeuthen, Germany 

$^i$\ KFKI, Budapest, Hungary 
}

  \centerline{\footnotesize presented by Zh.DZHILKIBAEV } 
  \centerline{\footnotesize {\it Institute for Nuclear Research, 60-the October Anniversary prospect 7a}} 
  \centerline{\footnotesize {\it Moscow 117312, Russia}} 
  \centerline{\footnotesize E-mail: djilkib@pcbai10.inr.ruhep.ru}

  \vspace*{0.9cm}
  \abstracts{
 We review the present status of the Baikal Neutrino Project.
The construction and performance of the large deep underwater
Cherenkov detector NT-200 with 192 PMTs , 
which is currently 
taking data in Lake Baikal, are described. Some results
from intermediate detector stages are
presented.
}   
  \normalsize\baselineskip=15pt
  \setcounter{footnote}{0}
  \renewcommand{\thefootnote}{\alph{footnote}}
  \section{Detector }
The deep underwater Cherenkov detector {\it NT-200},
the medium-term goal of the BAIKAL collaboration \cite{Project,APP,APP2},
was put into operation 
at April 6th, 1998. {\it NT-200}
is deployed in Lake Baikal, Siberia, 
\mbox{3.6 km} from shore at a depth of \mbox{1.1 km}. 
The detector comprises 192 optical modules (OM) at 8
vertical strings, see Fig.1.

The OMs are grouped in pairs along the strings. They contain 
37-cm diameter {\it QUASAR} PMTs which have been developed
specially for our project \cite{Project,APP,OM2}. The two PMTs of a
pair are switched in coincidence in order to suppress background
from bioluminescence and PMT noise. A pair defines a {\it channel}. 

All OMs face downward, with the exception of the OMs of the
second and eleventh layers, which look upward. The distance
between downward oriented layers is 6.25\,m, 
the distance between layers facing to each other 
(layers 1/2 and 10/11) is
7.5\,m, the distance between back-to-back layers (2/3 and 11/12)
is 5.0\,m.

A {\it muon-trigger}
is formed by the requirement of \mbox{$\geq N$ {\it hits}}
(with {\it hit} referring to a channel) within \mbox{500 ns}.
$N$ is typically set to 
\mbox{3 or 4.} For  such  events, amplitude and time of all fired
channels are digitized and sent to shore. 
A separate {\em monopole trigger} system searches for clusters of
sequential hits in individual channels which are
characteristic for the passage of slowly moving, bright
objects like GUT monopoles.

\begin{figure}
\centering
  \mbox{\epsfig{file=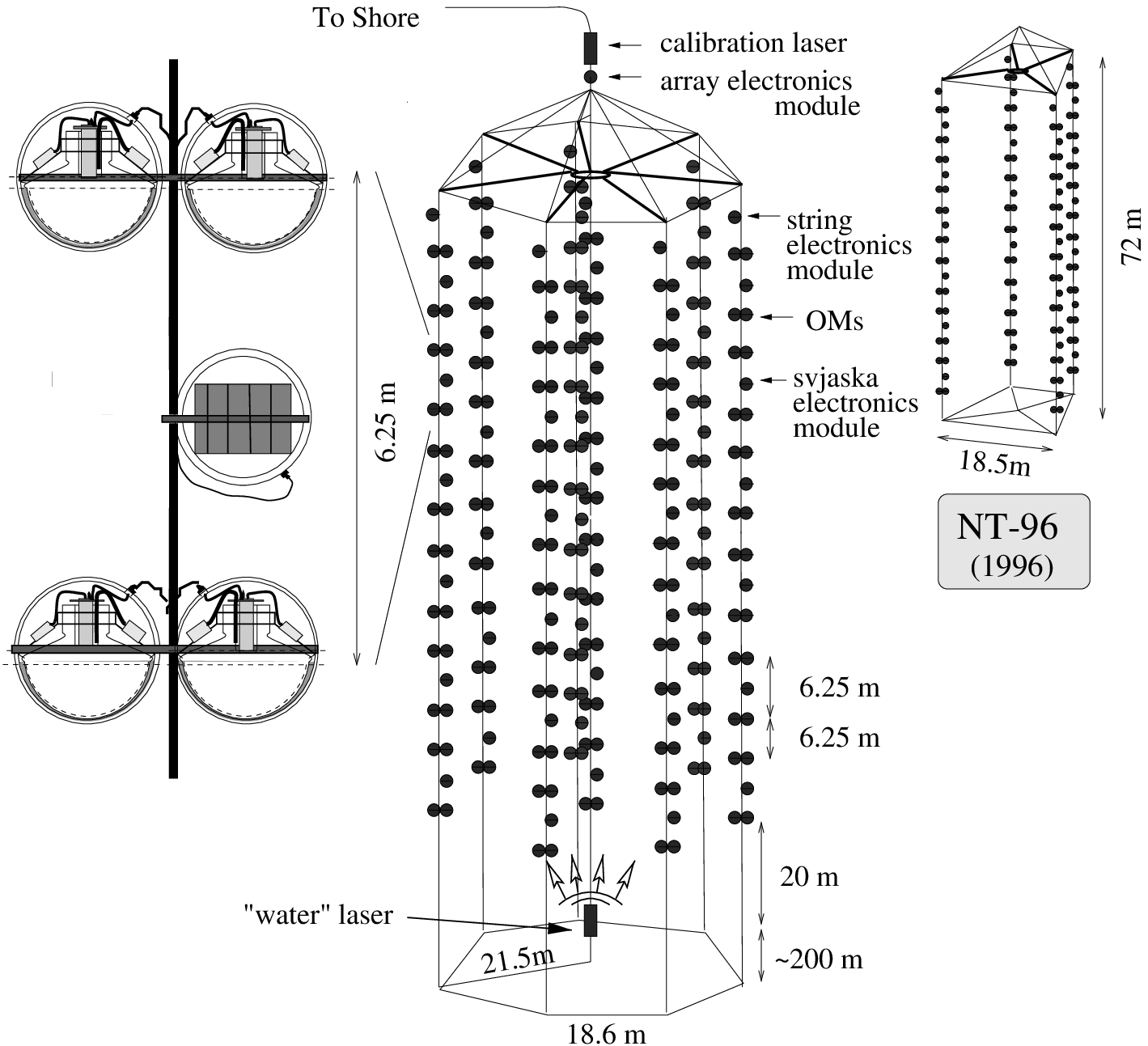,width=9.5cm}}
\fcaption{\small
Schematic view of the Baikal Telescope {\it NT-200}. The array
is time-calibrated by two nitrogen lasers. The one (fiber laser)
is mounted just above the array. Its light is guided via optical
fibers to each OM pair. The other (water laser) is arranged
20\,m below the array. Its light propagates directly through the
water.                  The expansion left-hand shows 2 pairs of
                  optical modules ("svjaska") with the svjaska
                  electronics module, which houses
                  parts of the read-out and control electronics.
Top right, the 1996 array {\it NT-96} is sketched.            
}
\end{figure}

In April 1993, the first part of {\it NT-200}, the detector {\it
NT-36} with 36 OMs at 3  strings, was put into operation 
and took data up to March 1995. A 72-OM array, {\it \mbox{NT-72}}, 
run in \mbox{1995-96}. In 1996 it
was replaced by the four-string array {\it NT-96}. 
{\it
  NT-144}, a six-string array with 144 OMs, was taking data
in \mbox{1997-98}.


Analysis of experimental data taken with intermediate arrays,
especially with {\it NT-36} and {\it NT-96}, 
proves the capability
of the Baikal neutrino telescope 
to investigate various
problems of
neutrino and muon physics. Below we present results which
illustrate the capability of the Baikal experiment to search for
atmospheric muons and neutrinos, neutrinos induced by
neutralino annihilation in the center of the Earth, magnetic
monopoles and showers produced by high energy neutrinos.

  \section{Atmospheric Muons}
Muon angular distributions as well as depth dependence of the
vertical flux obtained from data taken with {\it NT-36}
have been presented earlier \cite{APP}. 

Another example which confirms the efficiency of track reconstruction
procedure relates to the investigation of the shore ``shadow'' in muons with
{\it NT-96}. 
The Baikal Neutrino Telescope is placed  at a distance
of 3.6 km to the nearby shore of the lake. The opposite
shore is about 30 km away.
This asymmetry opens the possibility to investigate
the influence of the close shore to the azimuth distribution
under large zenith angles, where reconstruction for the
comparatively "thin" {\it NT-96} is most critical.
A sharp decrease of the
muon intensity at zenith
angles of 70$^0$-90$^0$  is expected. The comparison of the
experimental muon angular
distribution with MC calculations  gives us an estimation of
the accuracy of the
reconstruction error close to the horizontal direction.
Indeed, the {\it NT-96} data show a pronounced dip
of the muon flux in the direction of the shore and for
zenith angles larger than 70$^0$ --
in very good agreement with calculations which take into the
effect of the shore.

\begin{figure}
\centering
\mbox{\epsfig{file=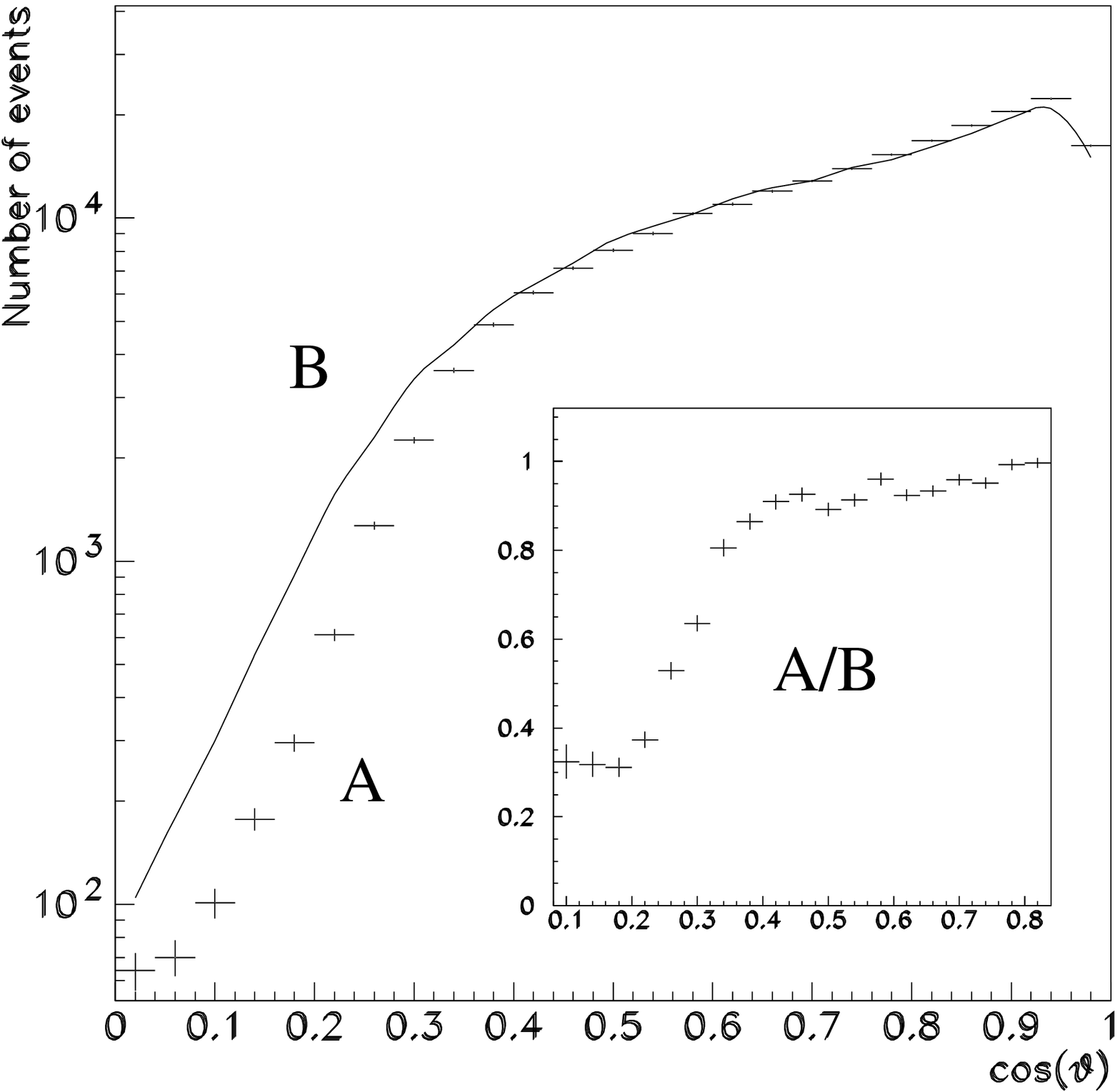,width=9.5cm}}
{\fcaption {
Atmospheric muons (vs zenith angle $\theta$) as it is 
measured in the direction to the nearest
point of the shore(A) and in opposite one(B) - ``open'' water. 
The small picture shows the ratio A to B.
}}
\end{figure}
  
  \section{Atmospheric Neutrinos}
The main results have been obtained with the first small 
detector {\it NT-36} - investigation of atmospheric 
muon flux, searching for
nearly vertically upward moving muons and searching for slowly moving
GUT monopoles  have been presented elsewhere \cite{APP,FRST_vert,GUT_monop}.
Below we present selected results obtained with
 {\it NT-96}.

\subsection{Identification of nearly vertically upward moving muons}    
Different to the standard analysis \cite{APP}, the method presented
in this section relies on the application of
a series of cuts which are tailored to the response
of the telescope to nearly vertically upward moving muons \cite{FRST_vert,INR_vert}.
The cuts remove muon events far away from the opposite
zenith as well as  background events which are mostly due to
pair and bremsstrahlung showers below the array and to naked downward
moving atmospheric muons with zenith angles close to the horizon
($\theta>60^{\circ}$). 
The candidates identified by the cuts are afterwards fitted in order to
determine the zenith angle.

We included all events with $\ge$4 hits along at least one
of all hit strings.
To this sample, a series of 6 cuts is applied. Firstly,
the time differences of hit channels along each individual  string
have to be compatible with a particle close to the opposite zenith (1). 
The event length should be large enough (2), the maximum recorded
amplitude  should not exceed a certain value (3) and
the center of gravity of hit channels should not be close to
the detector bottom (4). The latter two cuts reject
efficiently brems showers from downward muons.
Finally, also time differences of hits along {\it different}
strings have to correspond to a nearly vertical muon (5) and
the time difference between top and bottom hit in an event
has to be larger than a minimum value (6).

The effective area for muons moving close to opposite zenith
and fulfilling all cuts  exceeds $1000$ m$^2$. 

Within 70 days of effective data taking, $8.4 \cdot 10^7$ events
with the muon trigger $N_{hit} \ge 4$ have been selected. 

Table\,1  summarizes the number of events
from all 3 event samples (MC signal and background, and experiment)
which survive the subsequent cuts. 
After applying all cuts, four events were selected as neutrino
candidates, compared to 3.5 expected from MC. 
One of the four events
has 19 hit channels on four strings and was selected
as neutrino candidate by the standard analysis too. The zenith angular
distribution of these four neutrino candidates 
is shown in the inner box of Fig.3.

\begin{table}
\tcaption{
The expected number of atmospheric neutrino events and background
events, and the observed number of events after cuts{\it 1--6.}}
\small
\begin{center}
\begin{tabular} {||c|c|c|c|c|c|c||} \hline \hline
after cut {\cal N}$^o$  $\rightarrow$ & 1 & 2 & 3 & 4 & 5 & 6 \\ \hline 
atm. $\nu$, MC & 11.2 & 5.5 & 4.9 & 4.1 & 3.8 & 3.5 \\ \hline
background, MC & 7106 & 56 & 41 & 16 & 1.1 & 0.2 \\ \hline
experiment     & 8608 & 87 & 66 & 28 & 5 & 4 \\ \hline \hline
\end{tabular}
\end{center}
\end{table}

Regarding the  four detected events as being due to
atmospheric neutrinos, one can derive an 
the upper limit on the flux of muons from the center of the Earth   
due to annihilation of neutralinos - the favored candidate for
cold dark matter.

The limits on the excess muon flux obtained with underground 
experiments \cite{Bak,MACRO,Kam} and {\it NT-96} are shown in Table 2.
The limits obtained with {\it NT-96} 
are 4--7 times worse then the best underground limits since
the data collecting time of {\it NT-96} was only $\approx 70$ days. 
%

\begin{table}[h]
\label{limit}
\tcaption{90\% C.L. upper limits on the muon flux from the center of the Earth
for four regions of zenith angles obtained in different experiments
}
\small
\begin{center}
\begin{tabular}{||c|c|c|c|c||} \hline \hline
      &\multicolumn{4}{c|}{Flux limit       ($10^{-14} \cdot (cm^2 \; sec)^{-1})$} \\ \cline{2-5} 
Zenith   & {\it NT-96}  & {\it Baksan} & {\it MACRO}  & {\it Kam-de}  \\ 
angles   &$>10GeV$       &$>1GeV$        &$>1.5GeV$      &$>3GeV$ \\ \hline
$\geq 150^{\circ}$  & $11.0$      & $2.1$ & $2.67$ &$4.0$ \\ \hline
$\geq 155^{\circ}$  & $9.3 $      & $3.2$ & $2.14$ &$4.8$ \\ \hline
$\geq 160^{\circ}$  & $ 5.9-7.7 $ & $2.4$ & $1.72$ &$3.4$ \\ \hline
$\geq 165^{\circ}$  & $4.8$       & $1.6$ & $1.44$ &$3.3$ \\ \hline \hline
\end{tabular} 
\end{center}
\end{table}

This result, however, illustrates 
the capability of underwater experiments with respect
to the search for muons due to
neutralino annihilation in the center of the Earth.

\subsection{Selection of neutrino events over a large solid angle}
The signature of neutrino induced events is a muon crossing
the detector from below.
With the flux of downward muons exceeding that of
upward muons from atmospheric neutrino interactions by
 about 6 orders of magnitude, a careful reconstruction is of
prime importance.

In contrast to first stages of the detector 
({\it NT-36} \cite{FRST_vert}),
{\it NT-96} can be
considered as a real  neutrino telescope for a wide region in
zenith angle $\theta$.
After the reconstruction of all events with $\ge$ 9
hits at $\ge$
3 strings (trigger{\it 9/3}), quality cuts have been applied
in order to reject fake events.
Furthermore, in order to guarantee a minimum lever arm for track
fitting, events with a projection of the most distant channels on
the track ($Z_{dist}$) less than 35 meters have been rejected.
Due to the small transversal
dimensions of {\it NT-96}, this cut excludes zenith angles close to
the horizon.

The efficiency of the procedure has been tested with
a sample of $ 1.8 \cdot 10^6$ MC-generated atmospheric
muons, and with MC-generated upward muons due to atmospheric neutrinos.
It turns out that the signal to noise ratio is $ > 1$ for this sample.

The reconstructed angular distribution of \mbox{$2 \cdot 10^7$} events
taken with
{\it NT-96} in April/September 1996 -- after all cuts -- is 
shown in Fig.3.
%
%
\begin{figure}
\centering
  \mbox{\epsfig{file=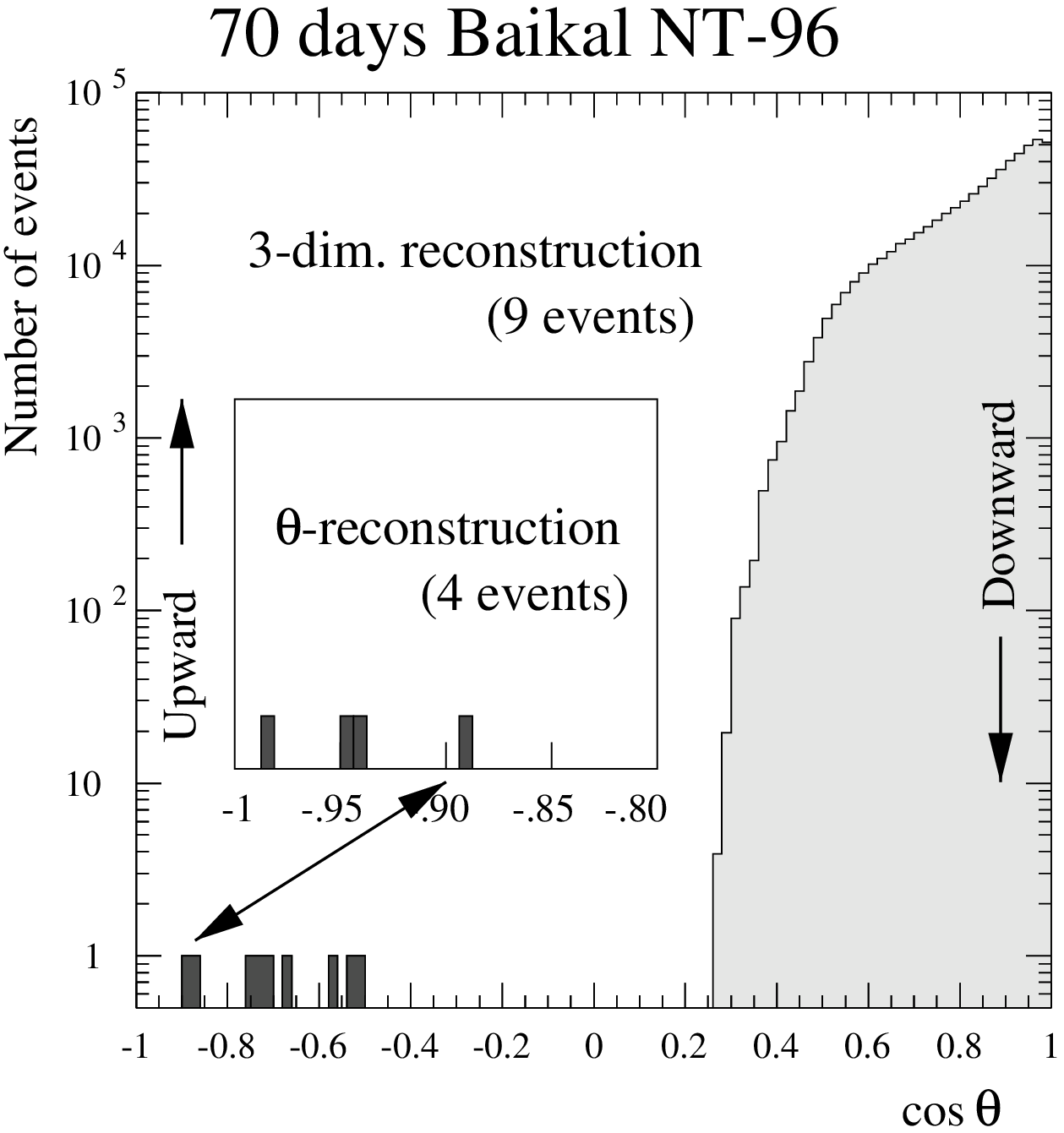,width=9.5cm}}
\fcaption{\small
Experimental angular distribution of events satisfying
trigger
{\it 9/3}, all final quality cuts and the limit on $Z_{dist}$ (see
text). The sub-picture shows the events selected by using the method 
described in subsection 3.1. The event found by both algorithms is
marked by the arrow.
}
\end{figure}

From 70 days of {\it \mbox{NT-96}} data,
12 neutrino candidates have been found. Nine of them have been
fully reconstructed. Three nearly upward vertical tracks
(see subsection 3.1)
hit only 2 strings and give a clear zenith angle but
ambiguities in the azimuth angle -- similar to the two events from
{\it NT-36} \cite{APP}. 
This is in  good agreement with MC expectations.

\section{Search for Fast Monopoles ($\beta > 0.75$)}
Fast bare monopoles with unit magnetic Dirac charge and velocities greater
than the Cherenkov threshold in water ($\beta = v/c > 0.75$) are
promising survey objects for underwater neutrino telescopes. 
For a given velocity $\beta$ the monopole  Cherenkov  radiation exceeds that
of a relativistic muon by a factor $(gn/e)^2=8.3\cdot10^3$ ($n=1.33$ -
index of refraction for water)
\cite{Fr,DA}.  
Therefore
fast monopoles with $\beta \ge 0.8$ can 
be detected up to distances 
$55$ m $\div$ $85$ m which corresponds to effective 
areas of (1--3)$\cdot 10^4$ m$^2$.

\begin{figure}
\centering
\mbox{\epsfig{file=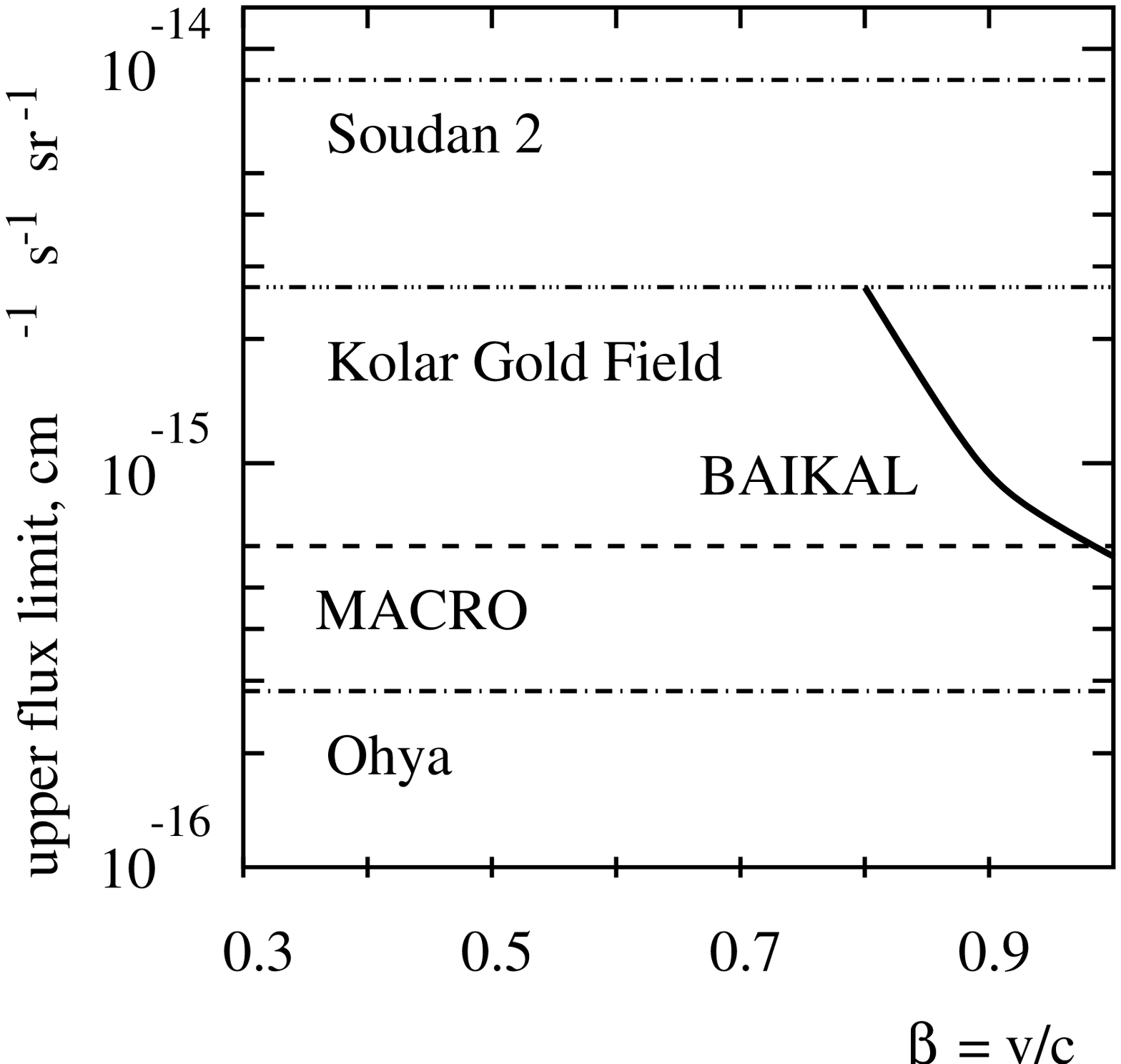,width=9.5cm}}
{\fcaption { 
The $90\%$ C.L. Baikal upper limit for an isotropic flux of
bare magnetic monopoles compared with other published limits.   
}}
\end{figure}

The natural way for fast monopole detection is based on the
selection of events
with high multiplicity of hits. 
In order to reduce  the background from
downward atmospheric muons we restrict 
ourself to monopoles coming from the lower hemisphere.

Two independent approaches have been used for selection of upward monopole 
candidates from the 70 days of {\it NT-96} data.
The first one is similar to the method which was applied
to upward moving muons (see subsection 3.1), with an additional cut 
$N_{hit}>25$ on the hit multiplicity.
The second one cuts on the value of space-time correlation, followed
by a cut $N_{hit}>35$ on the hit multiplicity. 
The upper limits on the monopole flux 
obtained with the two different methods coincide
within errors.  

The same type of analysis was applied to 
the data taken during $0.42$ years lifetime with the neutrino
telescope {\it NT-36} \cite{INR}.
 
The combined $90\%$ C.L. upper limit obtained by the Baikal
experiment for an isotropic flux of bare fast magnetic monopoles 
is shown in Fig.4, together with the best limits from
underground experiments Soudan2, KGF, MACRO and Ohya
\cite{Oh,MA,KGF,Sou} in Fig.4.

\section{Search for Very High Energy Electron Neutrinos}
In this section we present very preliminary
results with the aim to illustrate
the capability
of the Baikal Neutrino Telescope to search for extraterrestrial high energy
neutrinos from AGNs, GRBs and other sources. 

The idea used here to search for high energy electron
neutrinos ($E_{\nu} > 100$ TeV) is to detect the Cherenkov 
light emitted by the electromagnetic and (or) hadronic
particle cascade produced at the neutrino interaction
vertex in the sensitive volume of the neutrino telescope.
Earlier this idea has been used by DUMAND \cite{DUMAND} and \mbox{AMANDA \cite{AMANDA}}
to obtain upper limits on the diffuse
flux of high energy neutrinos.

In order to reduce the background from downward moving
atmospheric muons we restrict ourself to cascades
produced in a sensitive volume below the detector (see Fig.5)
and cause high multiplicity events in detector.
The trigger conditions for event selection are the same
as those which were used for fast monopole detection (see sec.4).

\begin{figure}
\centering
\mbox{\epsfig{file=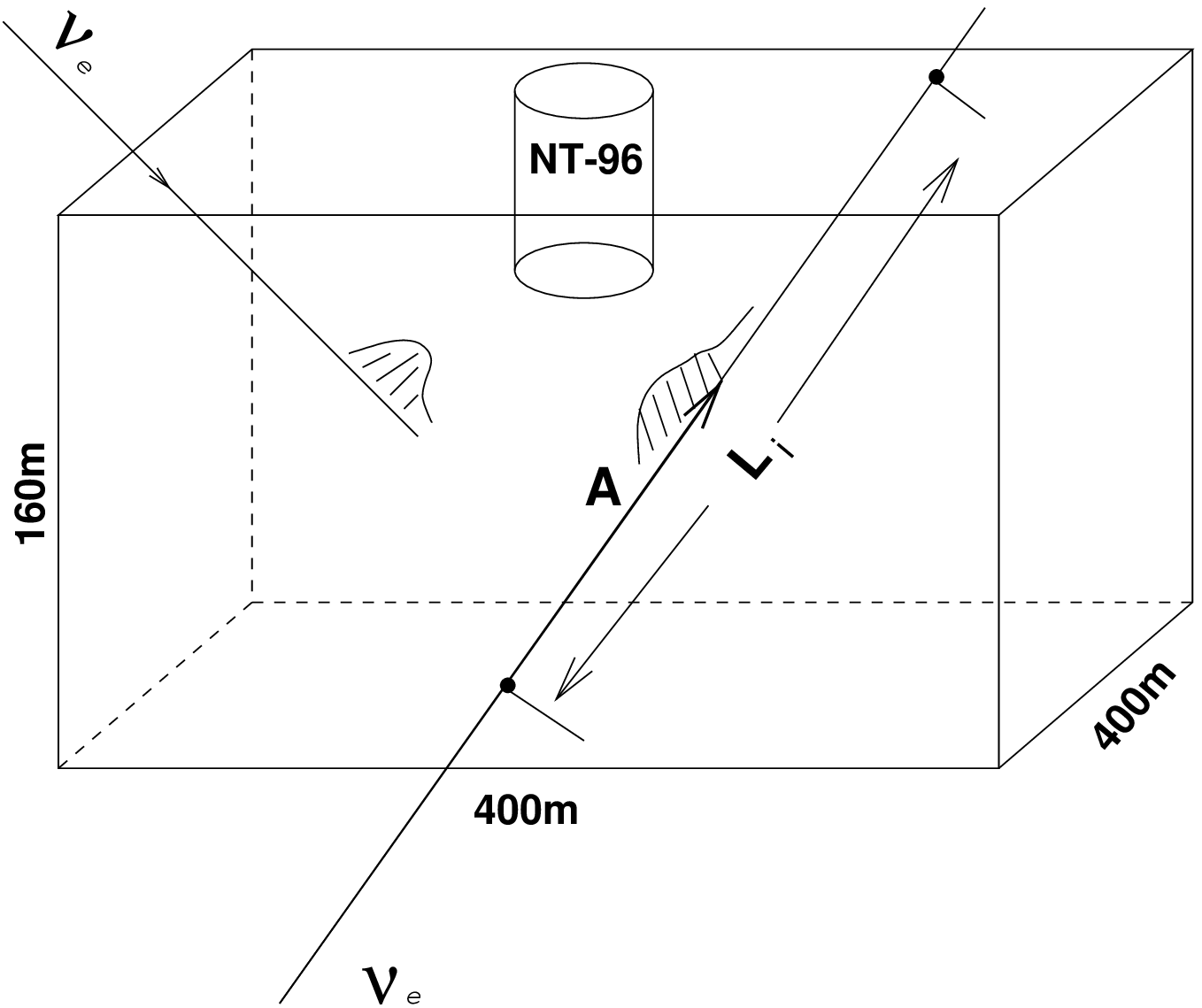,width=9.5cm}}
{\fcaption { 
The scheme illustrating the idea of neutrino induced
cascades detection in {\it NT-96}.
}}
\end{figure}

The effective volumes of {\it NT-96} 
averaged over neutrino directions
for detection of cascades
ith energy $E_{sh}$ are presented in Fig.6. The curves marked as
``DOWN'', ``UP'' and ``TOTAL'' correspond to 
effective volumes averaged over
lower and upper hemisphere and over all
directions. Also effective volumes of detectors SPS (DUMAND) 
and AMANDA-A are presented in Fig.6.

\begin{figure}
\centering
\mbox{\epsfig{file=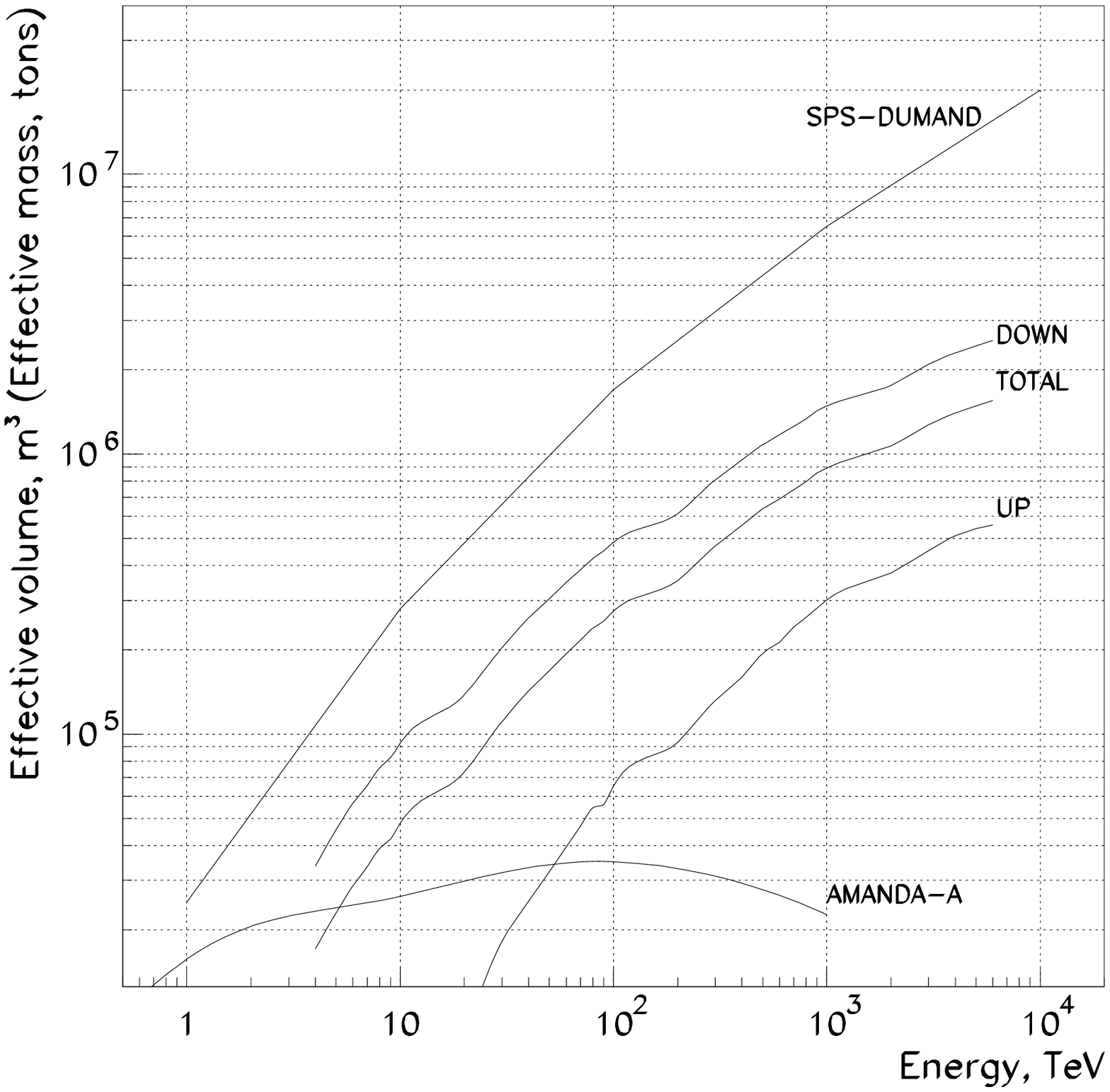,width=9.7cm}}
{\fcaption { 
Effective volumes of {\it NT-96}, SPS and AMANDA-A. ``DOWN'',
``UP'' and ``TOTAL'' correspond to effective volumes of {\it NT-96}
averaged over lower and upper hemisphere and over all directions of
neutrino velocity, respectively.
}}
\end{figure}

After analysis of 70 days of {\it NT-96}
data no evidence for any neutrino-induced cascades is found.

\subsection{The limit to the $\tilde{\nu_e}$ flux at the W resonance energy }
Although the neutrino-electron interactions
can generally be neglected with respect to neutrino-nucleon interactions
due to the small electron  mass,
the resonance cross section of $\tilde{\nu_{e}}e$ interaction at 6.3 PeV
is larger than the $\nu N$ cross section at any energy up to $10^{21}$ eV. 

The resonant cross section at 6.3 PeV for $\tilde{\nu_e}e$ scattering with 
a hadronic cascade in the final state:

\begin{equation}
\tilde{\nu_e} + e \rightarrow W^- \rightarrow hadrons
\end{equation}
 is $3.41 \times 10^{-31}$cm$^2$ \cite{Gandi}.
The cross section averaged over the energy range 

\begin{equation}
\Delta E=(M_w+2\Gamma_w)^2/2m_e - (M_w-2\Gamma_w)^2/2m_e, 
\end{equation}

$$
M_w=80.22 GeV, \, \, \, \Gamma_w=2.08 GeV
$$
is $\bar{\sigma}=1.12 \times  10^{-31}$cm$^2$. Eq.3 is used to
calculate the upper limit on the diffuse flux of $\tilde{\nu_e}$:

\begin{equation}
\frac{dF_{\tilde{\nu}}}{dE_{\tilde{\nu}}} \leq \frac{2.3}{\frac{10}{18}\rho N_A \bar{\sigma} T \Omega_{eff} V_{eff}\Delta E }.
\end{equation}
Here  T is the detector livetime (70 days), $\Omega_{eff}$ and $ V_{eff}$ are the average effective
solid angle and volume of the detector respectively. 

\pagebreak

The 90\% CL limit at the W resonance energy is:

\begin{equation}
\frac{dF_{\tilde{\nu}}}{dE_{\tilde{\nu}}} \leq 3.7 \times 10^{-18} cm^{-2}s^{-1}sr^{-1}GeV^{-1}.
\end{equation}
This limit lies between  limits obtained by SPS ($1.1 \times 10^{-18}$
cm$^{-2}$s$^{-1}$sr$^{-1}$GeV$^{-1}$) and EAS-TOP ($7.6 \times 10^{-18}$cm$^{-2}$s$^{-1}$sr$^{-1}$GeV$^{-1}$). 

\subsection{The limit to the $\nu_e + \tilde{\nu_e}$ flux}
For setting a limit to the $\nu_e + \tilde{\nu_e}$ flux
we have used the cross sections for $\nu_e$($\tilde{\nu_e}$) CC-interactions
with nucleons \cite{Gandi}

\begin{equation}
\nu_e(\tilde{\nu_e}) + N \stackrel{CC}{\rightarrow} e^-(e^+) + hadrons
\end{equation}
when all neutrino energy is transferred to the cascade.
The energy dependence of neutrino absorption in the Earth
has been taken into account.

Assuming the $F(E_{\nu})dE=A \delta (E_{\nu}-E)dE$ behavior of the differential neutrino
flux the 90\% CL limit has been obtained within the  $10^{13} \div 6 \times 10^{15}$eV, see Fig.7.

\begin{figure}
\centering
\mbox{\epsfig{file=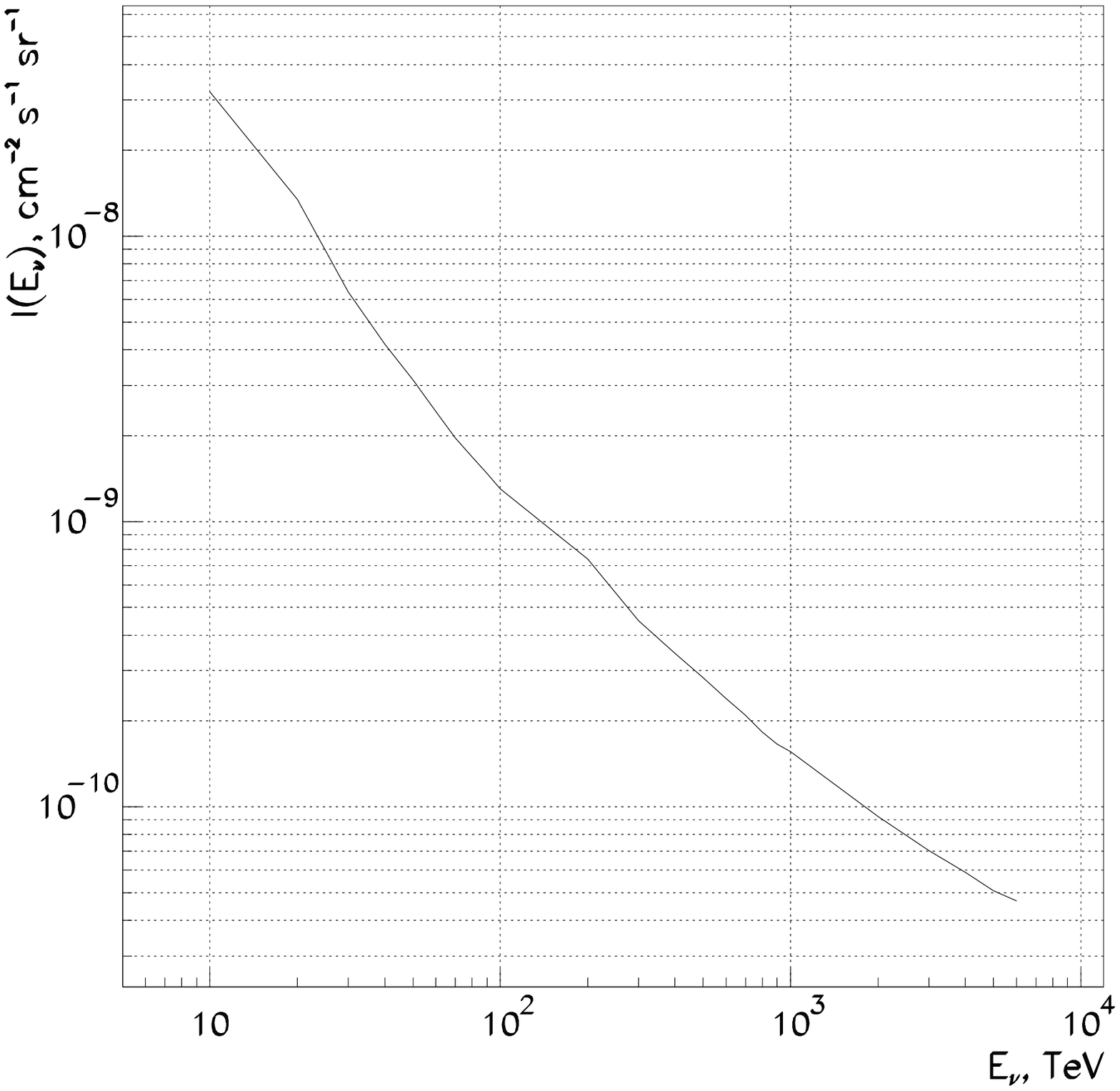,width=10.0cm}}
{\fcaption { 
The $90\%$ C.L. Baikal upper limit on the integral diffuse flux of
high energy $\nu_e + \tilde{\nu_e}$ neutrinos under the assumption of a $\delta (E_{\nu}-E)$
behavior of the neutrino flux.
}}
\end{figure}

To compare {\it NT-96} limit with those obtained by SPS and EAS-TOP \cite{EAS} 
we assume that a possible signal of 2.3 events originate in the energy interval from
$10^5$ to $10^6$ GeV with an $E^{-2}$ differential spectrum of neutrinos.

\begin{figure}
\centering
\mbox{\epsfig{file=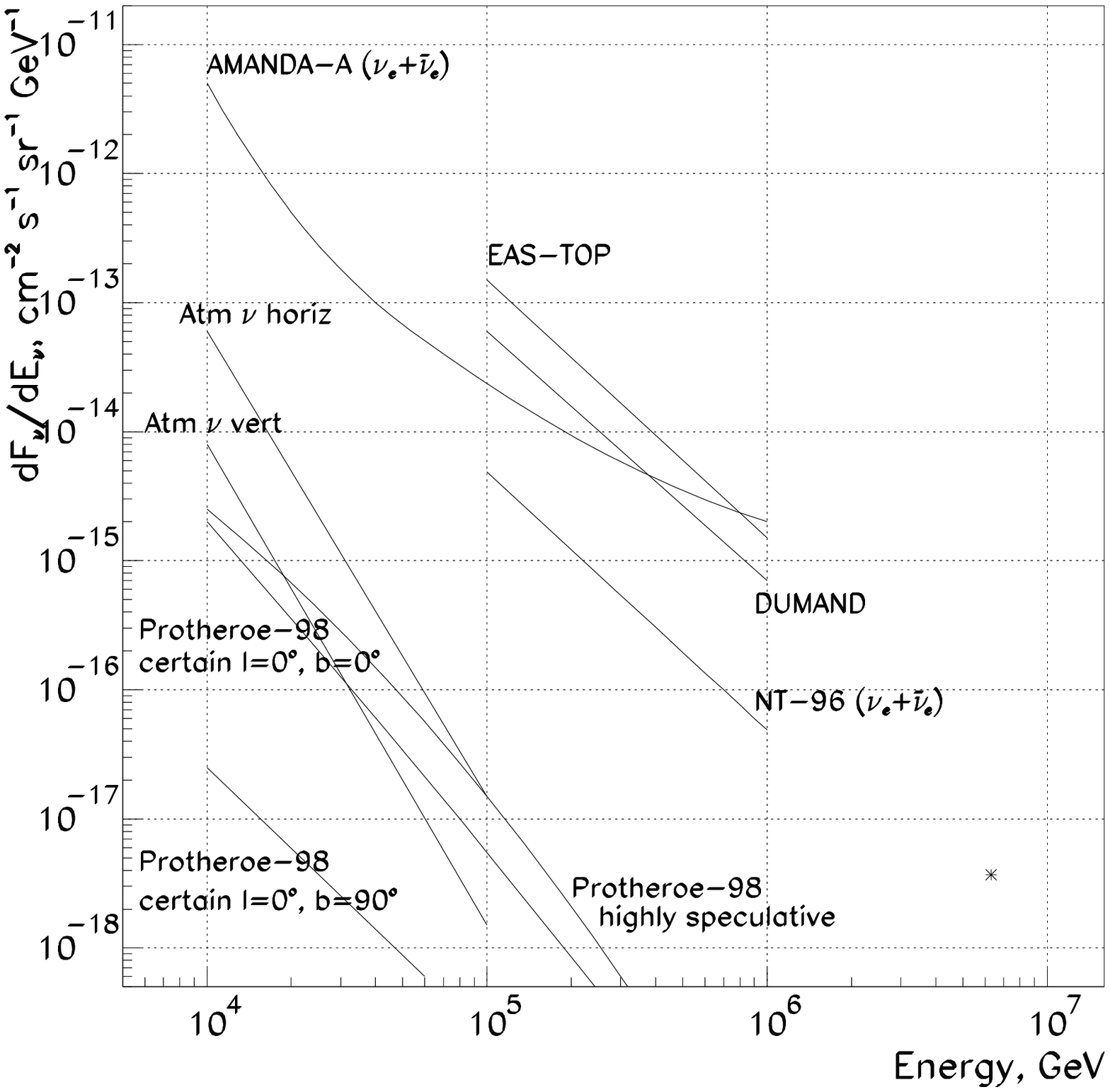,width=9.8cm,height=9.5cm}}
{\fcaption { 
The upper limits on the differential flux of high energy neutrinos obtained by 
different experiments and the resulting neutrino fluxes,
from a number of different models. 
Asterisk - the model-independed upper limit (4) on the $\tilde{\nu_e}$ flux  obtained at the W resonance
energy with {\it NT-96}.
}}
\end{figure}

This limit as well as limits obtained by other groups are shown
in Fig.8. Also, the resulting neutrino fluxes from a number of
different models [22] as well as backgrounds from atmospheric neutrinos [23]
are shown in Fig.8.

\section{Conclusions and Outlook}
The results obtained with intermediate detector stages
show the capability of Baikal Neutrino Telescope
to search for the wide variety of phenomena
in neutrino
astrophysics, cosmic ray physics and particle physics.

The first atmospheric neutrinos have
been identified. Also muon spectra have been measured,
and limits on the fluxes of magnetic monopoles as well as
of neutrinos from WIMP annihilation in the center of the Earth
have been derived. 

In the following years, {\it NT-200} will be operated as a
neutrino telescope with an effective area between 
1000 and 5000 m$^2$, depending on the energy and
will investigate atmospheric neutrino spectra above 10 GeV.

{\it NT-200} can be used to 
search for neutrinos from WIMP annihilation and for
magnetic monopoles. It will also be a unique
environmental laboratory to study water processes
in Lake Baikal.

Apart from its own goals, {\it NT-200} is regarded to be a 
prototype  
for the development a telescope of next generation 
with an effective area of
50,000 to 100,000 m$^2$.
The basic design of such a detector is under 
discussion at present.

\bigskip

{\it This work was supported by the Russian Ministry of Research,the German 
Ministry of Education and Research and the Russian Fund of Fundamental 
Research ( grants } \mbox{\sf 99-02-18373a}, \mbox{\sf 97-02-17935}, 
\mbox{\sf 99-02-31006}, \mbox{\sf 97-02-96589} and \mbox{\sf 97-05-96466}).

\end{document}